\documentclass[useAMS,usenatbib]{biom}
\def\bSig\mathbf{\Sigma}

\newcommand{\Perp}{\perp \! \! \! \perp}
\newcommand{\nPerp}{\not\!\perp\!\!\!\perp}
\providecommand{\tabularnewline}{\\}

\usepackage[figuresright]{rotating}
\usepackage{epstopdf}
\usepackage{amsmath}
\usepackage{amssymb}
\usepackage{graphicx}
\usepackage{esint}
\usepackage[authoryear]{natbib}
\usepackage{epstopdf}
\usepackage{amsmath}
\usepackage{amsfonts}
\usepackage{amssymb}
\usepackage{graphicx}

\title[Generalized Similarity U]{Generalized Similarity U:\\
A Non-parametric Test of Association Based on Similarity}

\author{Changshuai Wei$^{*}$\email{changshuai.wei@unthsc.edu} \\
	   Department of
Biostatistics and Epidemiology, University of North Texas Health Science
Center,\\
Fort Worth, TX 76107, USA
	   \and 
	   Qing Lu$^{*}$\email{qlu@epi.msu.edu}\\
	   Department of Epidemiology and Biostatistics, Michigan State University,
	   East Lansing, MI, 48824, USA
	   }

\begin{document}

%\date{{\it Received October} 2004. {\it Revised February} 2005.\newline  {\it Accepted March} 2005.}

\pagerange{\pageref{firstpage}--\pageref{lastpage}} \pubyear{xxxx}

\volume{xx}
\artmonth{xx}
\doi{xxxxxxx/xxxxxxxxxxxxxx}

%  This label and the label ``lastpage'' are used by the \pagerange
%  command above to give the page range for the article

\label{firstpage}

%  pub the summary here

\begin{abstract}
\\
\textbf{Motivation:} Second generation sequencing technologies are being increasingly used for genetic
association studies, where the main research interest is to identify sets of genetic
variants that contribute to various phenotype. The phenotype can be univariate disease status, multivariate responses and even high-dimensional outcomes. Considering the genotype and phenotype as two complex objects, this also poses a general statistical problem of testing association between complex objects.\\
\textbf{Results:} We here proposed
a similarity-based test, generalized similarity U (GSU), that can test the association between complex objects. We first studied the theoretical properties of the test in a general setting and then focused on the application of the test to sequencing association studies. Based on theoretical analysis, we proposed to use Laplacian kernel based similarity for GSU to boost power and enhance robustness. Through simulation, we found that GSU did have advantages over existing methods in terms of power
and robustness. We further performed a whole genome sequencing (WGS) scan for Alzherimer's Disease Neuroimaging Initiative (ADNI) data, identifying three genes, \textit{APOE}, \textit{APOC1} and \textit{TOMM40}, associated with imaging phenotype.\\
\textbf{Availability:} We developed a C++ package for analysis of whole genome sequencing data using GSU. The source codes can be downloaded at https://github.com/changshuaiwei/gsu. \\
\textbf{Contact:} weichangshuai@gmail.com\\

\end{abstract}

%
%  Please place your key words in alphabetical order, separated
%  by semicolons, with the first letter of the first word capitalized,
%  and a period at the end of the list.
%


\maketitle

\section{Introduction}

The ongoing sequencing studies allowed researchers to comprehensively investigate the role of a deep catalog of human genome variations in complex diseases\citep{cirulli2010}. Although these studies hold great promise for uncovering novel disease-associated variants, the massive sequencing data bring tremendous computational and statistical challenges to data analysis. Sequencing data is characterized with high-dimensionality and sparsity, where a large portion of genetic variants are rare variants with minor allele frequency (MAF) smaller than 5\%. Even with a large effect size,
a rare variant is hard to detect because of its low MAF. Moreover,
the massive number of rare variants raises computational burden
and multiple comparison issue. 

The common strategy is to perform a joint association test,
namely, testing the joint effect of a set of single nucleotide variants
(SNVs) on a genomic region, a functional unit (e.g., a gene) or a functional
pathway. By combining multiple SNVs,
the association information is aggregated and and the number of tests
is greatly reduced. Among these approaches, methods based on variance component score tests (VCscore) are widely used \citep{lin1997,Wu2011}. The methods considered the effects of the multiple variants as a random effect, and then test the effect by testing the variance component under the framework of the linear mixed model or the generalized linear mixed model.

There are also increasing interests in studying shared genetic contribution to multivariate phenotype. The multivariate phenotype can be multiple measurements
evaluating different aspects of a disease, which better reflect
the underlying biological mechanism of the disease. It can also be multiple disease
phenotypes that used for studying co-morbid genes or pleiotropic gene\citep{Dick2008}. A few methods can test the association of SNV-set with multivariate phenotype, yet, most of the current methods can not handle multivariate phenotype when the outcome variables are of different types (e.g., some variables are binary while others are continuous). Besides conventional multivariate phenotype, modern data types, such as shapes, images and trees, are emerging in biomedical researches. These complex objects are difficult to be integrated in traditional statistical frameworks, whose primary interests are variables in vector spaces(e.g. continuous, ordinal and categorical variables). Yet, it is relatively easy to define distance metric or similarity metric for complex objects. As a consequence, many distance and similarity based methods have been proposed for modern data analysis. 

In this paper, we present a similarity-based test using U statistic, referred to as the Generalized Similarity U test (GSU). GSU can be used to test the association of high-dimensional and sparse predictors with univariate, multivariate or complex-object responses from sequencing association studies or other association studies. We first studied the theoretical properties of GSU in a general setting in Section 2, where we investigated the finite-sample properties and asymptotic properties of the test. In section 3, we then focused on the applications of GSU to genetic sequencing studies. Extensive simulation studies were conducted to evaluated the performance of GSU in section 4, followed by a whole genome sequencing data application in section 5.

%\enlargethispage{12pt}

\section{Generalized Similarity U}

\subsection{General Setting and Rational}
We start with a formal set-up. Let $(\Omega, \mathcal{A}, P)$ and $(\Psi, \mathcal{G}, d)$ be a probability space and a metric space respectively. Consider random elements $Y$ and $G$ taking values on metric space $(\Psi_Y, \mathcal{G}_Y, d_Y)$ and $(\Psi_G, \mathcal{G}_G, d_G)$ with distribution $P_Y$ and $P_G$ respectively. Here, the random elements can be random variables (e.g., $\Psi=\mathbb{R}$), random vectors (e.g., $\Psi=\mathbb{R}^p$), random matrix (e.g., $\Psi=\mathbb{R}^{p \times p}$), random graph (e.g., trees), or random objects (e.g., shapes). 

Let $y$ and $g$ denote the realization of the random response element $Y$ and random predictor element $G$. Given a sample of data $\{(y_i, g_i)\}_{1 \leq i \leq n}$, we are interested in testing the association of response $Y$ and predictor $G$. Since $Y$ and $G$ may not live in a natural vector space, it is not straight forward to construct a regression model, such as $E(Y|G)=f(G)$. However, it is easier to construct similarity measurement for pairs $\{(y_i,y_j)\}_{i\neq j}$ and $\{(g_i,g_j)\}_{i\neq j}$ with the distance metrics $d_Y$ and $d_G$. Intuitively,  if $Y$ and $G$ are associated, then high similarity between $g_i$ and $g_j$ should lead to high similarity between $y_i$ and $y_j$. 

The similarity measurement can be defined by a real-valued function quantifying the similarity between two elements. For example, we can define the similarity between $y_i$ and $y_j$ as $h(y_i,y_j)=e^{-d_Y(y_i,y_j)}$, so that the closer $y_i$ and $y_j$ are in the metric space, the more similar $y_i$ and $y_j$ are. Other possible transformations from distance to similarity include inverse transformation $h(y_i,y_j)=(d_Y(y_i,y_j)+\varepsilon)^{-1}$ (for some $\varepsilon>0$) and thresholding transformation $h(y_i,y_j) = 1_{\{d_Y(y_i,y_j)<\varepsilon\}}$ ( for some $\varepsilon>0$). Loosely speaking, any monotonically non-increasing function can be used to transform distance to similarity. 

Here, we list some examples of similarity measurements.

Example 1 (vector similarity): Let $y_i,y_j \in \mathbb{R}^p$. We can use Gaussian kernel ($h(y_i,y_j)=exp(-||y_i-y_j||_2^2)$) or cross-product kernel ($h(y_i,y_j)=<y_i,y_j>$) to measure similarity. Here, $<y_i,y_j>$ can be considered as a transformation from Euclidean distance $||\cdot||_2^2$, using the fact that $<y_i,y_j>= -1/2(||y_i-y_j||_2^2 - ||y_i||_2^2 -||y_j||_2^2)$.

Example 2 (graph similarity): Let $y=(V,E)$ be a graph with adjacency matrix $A$, where $V$ is the set of vertices and $E$ is the set of edges. For any two graphs  $y_i$ and $y_j$, we can construct a product graph $y_i\otimes y_j$, with adjacency matrix $A_i \otimes A_j$. The similarity between the two graphs can be calculated using random walk on the product graph, $h(y_i,y_j)=\sum_k w_k q_{\otimes}^TW_{\otimes}^kp_{\otimes}$ \citep{Vishwanathan2010}, where $k$ is the length of the random walk, $w_k$ is the weight for size-$k$ random walk, $p_{\otimes}$ is the initial probability for vertices on $y_i\otimes y_j$, $W_{\otimes}$ is the transition probability obtained from $A_i \otimes A_j$, and $q_{\otimes}$ is the stopping probability for vertices on $y_i\otimes y_j$. Beside random walk, graph similarity can also be calculated using graphlet and subtree pattern.

Example 3 (image similarity): Image similarity can be calculated from local features and global features of the images by using traditional computer vision techniques such as scale invariant feature transformation (SIFT) and histogram of gradients (HOG). Both SIFT and HOG are human designed feature extraction. With large data sets, we can use modern machine learning methods, such as deep neural network \citep{Lecun2015}, to automatize the feature extraction, and construct more meaningful image similarity from high level representation of image.

\subsection{A Motivating Model}
Given the predictor elements and the response elements for the subjects $i$ and $j$
,we denote their response similarity $S_{i,j}$ by,\[S_{i,j}=h(y_{i},y_{j}),\]
and denote their predictor similarity $K_{i,j}$ by,$K_{i,j}=f(g_{i},g_{j})$. The similarity measurements $h(\cdotp,\cdotp): \Psi_Y \times \Psi_Y \to \mathbb{R}$ and $f(\cdotp,\cdotp): \Psi_G \times \Psi_G \to \mathbb{R}$
can be of a general form as long as they satisfy the
finite second moment condition, i.e., $E(h^{2}(Y_{1},Y_{2}))<\infty$
and $E(f^{2}(G_{1},G_{2}))<\infty$, where $Y_1$ and $Y_2$ ($G_1$ and $G_2$) are independent identical copy of $Y$ ($G$). We center the response
similarity $\tilde{S}_{i,j} = \tilde{h}(y_{i},y_{j})$ by,
\begin{align}
\tilde{h}(y_{i},y_{j})= & h(y_{i},y_{j})-E(h(y_{i},Y_{j}))\nonumber\\
 & -E(h(Y_{i},y_{j}))+E(h(Y_{i},Y_{j})),\label{eq:Centering}
\end{align}
and center the predictor similarity, $\tilde{K}_{i,j}=\tilde{f}(g_{i},g_{j})$,
in the same manner. Based on the definition of the centered similarity, we can show that
$E(\tilde{f}(G_{i},G_{j}))=0$ and $E(\tilde{h}(Y_{i},Y_{j}))=0$
(Supplementary Appendix S1).  

We can investigate the relationship of the two similarities using a similarity regression model \citep{Elston2000,Tzeng2009}, \[
E(\tilde{S}_{r}|\tilde{K}_{r})=b\tilde{K}_{r}, \forall r \in \{(i,j),i < j\}.\]
Since the similarities have been centered, the regression has zero intercept. The association can then be evaluated by testing null hypothesis $b=0$, where $b$ can be estimated by,
$
\hat{b}={\sum_{i < j} \tilde{K}_{i,j}\tilde{S}_{i,j}}/{\sum_{i < j} (\tilde{K}_{i,j})^2}.
$
By the form of $\hat{b}$, testing $b=0$ is equivalent to testing the numerator $U_b=0$, where $U_b= \sum_{i < j} \tilde{K}_{i,j}\tilde{S}_{i,j}$. As we shall see soon, $U_b$ is in the same form as the generalized similarity U.

\subsection{Weighted U Statistic}

The generalized similarity U (GSU) is defined
as the summation of the centered response similarities weighted by
the centered predictor similarities,
\begin{equation}
U=\frac{1}{n(n-1)}\sum_{i\neq j}\tilde{K}_{i,j}\tilde{S}_{i,j},\label{eq:GSU}
\end{equation}
where $\tilde{K}_{i,j}$ is considered as the weight function
and $\tilde{S}_{i,j}$ is considered as the U kernel. In our definition
of GSU, the role of response similarity and predictor similarity are
interchangeable. In other words, we can also treat $\tilde{S}_{i,j}$
as the weight function and $\tilde{K}_{i,j}$ as the U kernel.

Under the null hypothesis,
when the predictor element $G$ is independent of response element $Y$ (i.e., $Y \Perp G $),
we have $
E(U)  =  \frac{1}{n(n-1)}\sum_{i\neq j}E(\tilde{f}(G_{i},G_{j}))E(\tilde{h}(Y_{i},Y_{j})) = 0$.
Under alternative hypothesis, when $Y$ is associated with $G$, we expect that the response similarity
is concordant with the predictor similarity. In other words, the positive
response similarities are weighted heavier and the negative response
similarities are weighted lighter, leading to a positive value of
U statistic. A statistical test can be formed to test the association,
and p-value can be calculated by $P(U>U_{obs})$ under null hypothesis, where $U_{obs}$
is the observed value of $U$.

Define a population parameter $\mu_U$ as $$\mu_U=E(\tilde{f}(G_{1},G_{2})\tilde{h}(Y_{1},Y_{2})).$$ 
It is easy to show that GSU is an unbiased estimator
of $\mu_U$, i.e., $E(U)=\mu_U$.
In addition, knowing that $\mu_U=E(\tilde{f}(G_{1},G_{2})\tilde{h}(Y_{1},Y_{2}))-E(\tilde{f}(G_{1},G_{2}))E(\tilde{h}(Y_{1},Y_{2})),$
we can consider $\mu_U$ as a population covariance. In this sense, a scale invariant ``correlation",  $U_{\gamma}$, can be calculated,
$
U_{\gamma}={\sum_{i \neq j} \tilde{K}_{i,j}\tilde{S}_{i,j}}/{({ \sum_{i \neq j} (\tilde{K}_{i,j})^2  \sum_{i \neq j} (\tilde{S}_{i,j})^2})^{1/2}},
$
as an indicator of strength of association.

\subsection{Strongly Positive Definite Similarity}
We have already shown that $Y \Perp G \Rightarrow \mu_U=0 $, which ensures the correct type I error. It is of interest now whether $\mu_U=0 \Rightarrow Y \Perp G$, so that we can control type II error (i.e., improve power) and reject null hypothesis whenever $Y \nPerp G$. The establishment of $\mu_U=0 \Rightarrow Y \Perp G$ needs additional assumptions on the similarity measurements and metric spaces. For the completeness, we first introduce several preliminaries.

Define a ``kernel'' as a real symmetric function $h: \Psi \times \Psi \to \mathbb{R}$. A kernel is called positive definite if $\sum_{i,j}^n c_ic_jh(y_i,y_j) \geq 0$, $\forall$ $c_i,c_j \in \mathbb{R}$ and $\forall$ $y_i,y_j \in \Psi$. A kernel is called negative definite if $\sum_{i,j}^n c_ic_jh(y_i,y_j) \leq 0$, $\forall$ $c_i,c_j \in \mathbb{R}, y_i,y_j \in \Psi$ and $\sum_i c_i=0$.

A positive definite kernel is called strictly positive definite when the equality $\sum_{i,j}^n c_ic_jh(y_i,y_j) = 0$ implies $c_i=0$ $\forall i$. The kernel function here can be used to define similarity measurement. To consider $\mu_U=0 \Rightarrow Y \Perp G$, however, we need the kernel function to exhibit a property of ``strong'' positive definiteness in the integral form. Using similar notions of \cite{Rachev2013}, we define a strongly positive definite kernel as follows.

\textbf{Definition 1}:\textit{ Let Q be a finite positive measure on $(\Psi,\mathcal{G},d)$ and $q$ be a function integrable with respect to Q. We say $h$ is strongly positive definite if it is positive definite and the equality $\int_{\Psi}\int_{\Psi} h(x,y)q(x)q(y)dQ(x)dQ(y)=0$ implies $q=0$ a.e. $\forall$ Q.}

Let $\vartheta$ be a finite signed measure dominated by $Q$ s.t. $d\vartheta=qdQ$. For strongly positive definite kernel $h$, the equality $\int_{\Psi}\int_{\Psi} h(x,y)d\vartheta(x)d\vartheta(y)=0$ implies $\vartheta=0$ $\forall$ $\vartheta$. Now let $\vartheta=P_{GY}-P_GP_Y$ be a measure on $\Psi_G \times \Psi_Y$, we can show (in Supplementary Appendix S1) that 
\[
\mu_U=\int\int f(g_1,g_2)h(y_1,y_2)d \vartheta(g_1,y_1) d\vartheta(g_2,y_2)
\]
If the tensor product kernel $(f \otimes h)\big((g_1,y_1),(g_2,y_2)\big)=f(g_1,g_2)h(y_1,y_2)$ is strongly positive definite, then $\mu_U=0$ implies $\vartheta=0$ (i.e., $\mu_U=0 \Rightarrow Y \Perp G$). In fact, we can show $\mu_U=0 \Rightarrow Y \Perp G$ as long as $f$ and $g$ are both strongly positive definite. 

\textbf{Theorem 2}: \textit{Assume both $f(\cdot,\cdot)$ and $h(\cdot, \cdot)$ are strongly positive definite. Let $\tilde{h}(Y_{1},Y_{2})$ and $\tilde{f}(G_{1},G_{2})$
be the centered similarities as defined in (\ref{eq:Centering}). Define
$\mu_U=E(\tilde{f}(G_{i},G_{j})\tilde{h}(Y_{i},Y_{j}))$.
Then, $\mu_U=0 \Leftrightarrow Y \Perp G $.}

The proof is given in Appendix A by employing measures embedding into the reproducing kernel Hilbert space. Many popular kernels such as radial basis kernel $h(y_1,y_2)=exp(-||y_1-y_2||_q)$ ($0<q<2$) are strongly positive definite kernel on $\mathbb{R}^p$ \citep{Sriperumbudur2010}. However, the cross product kernel $h(y_1,y_2)=<y_i,y_j>$ is not strongly positive definite on  $\mathbb{R}^p$, by observing that $\int\int <y_1,y_2> d\vartheta(y_1)d\vartheta(y_2)=0 \Leftrightarrow \int y d\vartheta(y)=0 \nRightarrow \vartheta=0$.

\subsection{Asymptotic Test}
For high dimensional data, it is computationally expensive to calculate p-values $P(U>U_{obs})$ using permutation. Here, we derive the asymptotic distribution of GSU under null hypothesis.

By considering the predictor similarity as the weight function and the
response similarity as the U kernel, GSU is a weighted U statistic \citep{Lindsay2008,Wei2016}.
More specifically, because its kernel satisfied $Var(E(\tilde{h}(Y_{1},Y_{2})|Y_{2}))=0$
(Supplementary Appendix S1), GSU is a degenerated weighted
U statistic. To derive the limiting distribution of GSU, we can decompose
the centered response similarity by,
$
\tilde{h}(y_{1},y_{2})=\sum_{s=1}^{\infty}\lambda_{s}\phi_{s}(y_{1})\phi_{s}(y_{2}),
$
where $\{\lambda_{s}\}$ and $\{\phi_{s}(\cdotp)\}$ are eigenvalues
and eigenfunctions of the U kennel $\tilde{h}(\cdotp,\cdotp)$, and
all the eigenfunctions are orthogonal, i.e., $\int\phi_{s}(y_{1})\phi_{s'}(y_{1})dF(y_{1})$ equals 0 if $s\neq s'$ and equals 1 if $s=s'$.
Similarly, we can decompose the centered predictor similarity by,
$
\tilde{f}(G_{i},G_{j})=\sum_{t=1}^{\infty}\eta_{t}\varphi_{t}(g_{1})\varphi_{t}(g_{2})
$.
We can then rewrite the GSU as,
\begin{eqnarray*}
U &=& \frac{1}{n-1}\sum_{t=1}^{\infty}\sum_{s=1}^{\infty}\left(\frac{1}{\sqrt{n}}\sum_{i=1}^{n}\eta_{t}^{\star}(G_{i})\phi_{s}^{\star}(Y_{i})\right)^{2} \\
& & -\frac{1}{n-1}\sum_{t=1}^{\infty}\sum_{s=1}^{\infty}\frac{1}{n}\sum_{i=1}^{n}\left(\eta_{t}^{\star}(G_{i})\phi_{s}^{\star}(Y_{i})\right)^{2},
\end{eqnarray*}
where $\varphi_{t}^{\star}(G_{i})=\eta_{t}^{0.5}\varphi_{t}(G_{i})$
and $\phi_{s}^{\star}(Y_{i})=\lambda_{s}^{0.5}\phi_{s}(Y_{i})$. Using the form above, we can show that the limiting distribution of
GSU is a weighted sum of independent chi-square random variables.
This is the result of theorem 3 below, which is proved in Appendix B.

\textbf{Theorem 3}: \textit{Assume $E(h(Y,Y))< \infty$, $E(f(G,G)) < \infty$,
and $Y\Perp G$. Let $\tilde{h}(Y_{1},Y_{2})$ and $\tilde{f}(G_{1},G_{2})$
be the centered similarities as defined in (\ref{eq:Centering}). Define
U as $U=\frac{1}{n(n-1)}\sum_{i\neq j}\tilde{f}(G_{i},G_{j})\tilde{h}(Y_{i},Y_{j})$.
Then, $nU\xrightarrow{D}\sum_{t=1}^{\infty}\eta_{t}\sum_{s=1}^{\infty}\lambda_{s}(\chi_{st}^{2}-1)$,
where $\{\chi_{st}^{2}\}$ are independent chi-square random variables
with 1 degree of freedom.}

Using the similar techniques, we can show that a weighted V statistic in the following form,
$
V = \frac{1}{n^2} \sum_{i, j}\tilde{f}(G_{i},G_{j})\tilde{h}(Y_{i},Y_{j}),
$
also converges to a weighted sum of chi-squared variables, i.e.,
$
nV \xrightarrow{D} \sum_{t=1}^{\infty}\eta_{t}\sum_{s=1}^{\infty}\lambda_{s}\chi_{st}^{2}.
$

\subsection{Power and Sample Size}

In this subsection, we derive the asymptotic distribution of GSU under the
alternative hypothesis, and provide asymptotic power and sample size calculations
for association analysis.

Denote $\zeta_{1}=Var(\tilde{f}(G_{1},G_{2})\tilde{h}(Y_{1},Y_{2})|(G_{2},Y_{2}))$. Assume under the alternative hypothesis that $\mu_U>0$ and
$\zeta_{1}>0$. 
Using the Hoeffding projection, we can show that GSU asymptotically
follows a normal distribution, with mean $\mu_U$ and variance $4\zeta_{1}/n$.
This is the result of Theorem 4, which is proved in Supplementary Appendix S4.

\textbf{Theorem 4}: \textit{Let $\tilde{h}(Y_{1},Y_{2})$ and $\tilde{f}(G_{1},G_{2})$
be the centered similarities as defined in (\ref{eq:Centering}).
Suppose Y is associated with G, and the following conditions are
satisfied: $E(\tilde{f}(G_{1},G_{2})\tilde{h}(Y_{1},Y_{2}))=\mu_U>0$,
$Var(\tilde{f}(G_{1},G_{2})\tilde{h}(Y_{1},Y_{2}))=\zeta_{0}<\infty$,
and $Var(\tilde{f}(G_{1},G_{2})\tilde{h}(Y_{1},Y_{2})|(G_{2},Y_{2}))=\zeta_{1}>0$.
Define $U$ as $U=\frac{1}{n(n-1)}\sum_{i\neq j}\tilde{f}(G_{i},G_{j})\tilde{h}(Y_{i},Y_{j})$.
Then, $\sqrt{n}(U-\mu_U)\xrightarrow{D}N(0,4\zeta_{1})$.}

The power of GSU at the significance level $\alpha$ can be calculated
by,
$
P\{nU >q_{1-\alpha}\} 
= \Phi(\frac{n\mu_U-q_{1-\alpha}}{2\sqrt{n\zeta_{1}}}),
$
where $q_{1-\alpha}$ is the $1-\alpha$ quantile for $\sum_{t=1}^{\infty}\eta_{t}\sum_{s=1}^{\infty}\lambda_{s}(\chi_{st}^{2}-1)$
and $\Phi(\cdotp)$ is the CDF of a standard normal distribution.
The sample size required to achieve power $\beta$ can be calculated
by solving $\Phi(\frac{n\mu_U-q_{1-\alpha}}{2\sqrt{n\zeta_{1}}})\geq\beta$.
By denoting $Z_{\beta}$ as the $\beta$ quantile for a standard normal
distribution, the required sample size is given by,
$
n= \min_{n\in N}\left\{ n: \: n\geq {(Z_{\beta}\sqrt{\zeta_{1}}+({Z_{\beta}^{2}\zeta_{1}+\mu_U q_{1-\alpha}})^{1/2})^{2}}/{\mu_{U}^{2}}\right\} .
$
\section{Generalized Similarity U for Sequencing Association}

\subsection{Settings for Sequencing Data Analysis}

In a sequencing association study, the response element is called phenotype and the predictor element is called genotype. Common forms of phenotype and genotype are scalars or vectors. Suppose that $n$ subjects are sequenced in a study, where we are
interested in testing the association of L phenotypic variables ($y_{i,l}$, $1\leq i\leq n$,
$1\leq l\leq L$) with M genetic variants ($g_{i,m}$, $1\leq i\leq n$,
$1\leq m\leq M$). For each subject $i$, we observe a phenotype vector
$y_{i}$ ( $y_{i}=(y_{i,1},y_{i,2},\cdots,y_{i,L})$ ) and a genotype vector $g_{i}$ ( $g_{i}=(g_{i,1},g_{i,2},\cdots,g_{i,M})$).
In the special case when $L=1$ (or $M=1$), it is simplified to a
univariate analysis (or a single-locus analysis). When $L>1$ (or $M>1$), it
extends to a multivariate analysis (or a multi-locus analysis). Here, we allow multiple
phenotypes to be of different types (e.g., continuous or categorical),
and do not assume any distribution of phenotypes. The number of genetic
variants $M$ and the number of phenotypes $L$ can be larger than
the sample size. For example, the genetic data can be sequencing data (high
dimensional genotype) and the phenotype data can be imaging data (high dimensional phenotype).

\subsection{Similarity Measurement}

The choices for the phenotype similarity $h(\cdotp,\cdotp)$ and the
genetic similarity $f(\cdotp,\cdotp)$ are flexible. According to
different types of genetic variants and the purpose of the analysis,
we can choose different types of phenotype similarities and genetic
similarities.

For phenotype similarity, one popular approach is to use a cross product kernel, i.e., $h(y_i,y_j)=<y_i,y_j> $ \citep{Tzeng2009}. Yet, as discussed in previous theoretical analysis, cross product kernel may not fit for robust association analysis. Here, we propose a similarity measurement for both
categorical and continuous phenotype using radial basis kernel with L1 norm (Laplacian Kernel),
\[
S_{i,j}^{LK}=exp(-\sum_{l=1}^{L}\omega_{l}|y_{i,l}-y_{j,l}|),
\]
where $\omega_{l}$ represents the weight for the $l$-th phenotypes given based on prior knowledge. If there is no prior
knowledge, we can use an equal weight, $\omega_{l}=1/L$. The Laplacian Kernel (LK) based
phenotype similarity can be modified to take the correlation
among the phenotypes into account,
$
S_{i,j}^{LK}=exp\left(-\frac{1}{L}d_{ij}^{T}\Gamma d_{ij}\right),
$
where $d_{ij}=(|y_{i1}-y_{j1}|^{0.5},\cdots,|y_{iL}-y_{jL}|^{0.5})^{T}$. $\Gamma$ can be chosen
to reflect the correlations among the phenotypes. For example, we
can define $\Gamma$ as,
$
\Gamma=(\frac{1}{n}\sum_{i=1}^{n}y_{i}y_{i}^{T})^{-0.5}.
$

For the categorical SNVs data, the popular way of measuring genetic similarity is to use IBS function
or the weighted IBS function\citep{Lynch1999}.
Assuming the genetic variants ($g_{i,m}$, $1\leq i\leq n$, $1\leq m\leq M$)
are coded as 0, 1 and 2 for AA, Aa and aa respectively, the IBS-based genetic similarity is defined as,
$
K_{i,j}^{IBS}=\frac{1}{2M}\sum_{m=1}^{M}2-|g_{i,m}-g_{j,m}|.
$
Alternatively, the weighted-IBS (wIBS) genetic similarity can be defined
to emphasize the effects of rare variants,
$
K_{i,j}^{wIBS}=\sum_{m=1}^{M}{w_{m}(2-|g_{i,m}-g_{j,m}|)}/({2\Upsilon}),
$
where $w_{m}$ represents the weight for the $m$-th SNV in the SNV-set,
and $\Upsilon$ is a scaling constant, defined as $\Upsilon=\sum_{m=1}^{M}w_{m}$.
$w_{m}$ is usually defined as a function of minor allele frequency (MAF, denoted as $\gamma_{m}$). For
example, the weight $w_{m}$ can be calculated using inverse variance, i.e., $w_{m}=1/\sqrt{\gamma_{m}(1-\gamma_{m})}$. However, IBS-based similarity can not be used for other genetic data, such as copy number variation (count) or expression data (continuous). Here, we propose a unified LK-based genetic similarity by generalizing wIBS,
\[
K_{i,j}^{LK}=exp\big( -\sum_{m=1}^{M}\frac{w_{m}|g_{i,m}-g_{j,m}|}{\Upsilon} \big),
\]
where $g_{i,m}$ can be categorical, count or continuous variables, and $w_{m}$ can be calculated as function of variance $\sigma_m^2$ of $g_m$, i.e., $w_m=1/\sigma_m$.

Thus, we defined a unified measurement for genetic similarity and phenotype similarity with Laplacian kernel $exp(-|\cdot-\cdot|)$. Since laplacian kernel is strongly positive definite, we know that (from Theorem 2) the corresponding GSU has the property $\mu_U=0 \Leftrightarrow G\Perp Y$, so that it can control type II error for detection of any types of association. Since Laplacian kernel is bounded similarity measurement, i.e., $0 \leq h(\cdot,\cdot) \leq 1$ and $0 \leq f(\cdot,\cdot) \leq 1$, we know the regularity conditions in Theorem 3 is satisfied and the asymptotic test for corresponding GSU is robust against distribution assumptions (for large sample size).

\subsection{Computation and Covariates Adjustment}

Let $S=\{S_{i,j}\}_{n\times n}$ and $K=\{K_{i,j}\}_{n\times n}$
be the matrix form of the phenotype similarity and genetic similarity,
the centered similarity matrices $\tilde{S}$ and $\tilde{K}$ can
be obtained by,
$
\tilde{S}=(I-J)S(I-J),
$
and
$
\tilde{K}=(I-J)K(I-J),
$
where $I$ is an n-by-n identity matrix, and $J$ is an n-by-n matrix
where all elements are $1/n$ (Supplementary Appendix S5).
Then GSU can be expressed as,
$
U=\frac{1}{n(n-1)}\sum_{i\neq j}\tilde{K}_{i,j}\tilde{S}_{i,j}.
$
In this form, $U$ can be viewed as a sum of the element-wise product
of the two matrices, $\tilde{K}_{0}$ and $\tilde{S}_{0}$, which
are obtained by assigning 0 to the diagonal elements of matrices $\tilde{K}$
and $\tilde{S}$.

To allow for covariates adjustment, we can perform two sided projection on the zero-diagonal centered similarity matrices, $\tilde{K}_{0}$ and $\tilde{S}_{0}$. Suppose that there are $P$ covariates that need to be adjusted. Let $X=\{x_{i,p}\}_{n \times P}$ represents the covariate matrix, we can  calculate the covariate centered similarity matrices by (Supplementary Appendix S6),
$
\hat{S}=(I-X(X^TX)^{-1}X^T)\tilde{S}_0(I-X(X^TX)^{-1}X^T),
$
and
$
\hat{K}=(I-X(X^TX)^{-1}X^T)\tilde{K}_0(I-X(X^TX)^{-1}X^T).
$   
The covariate adjusted GSU can be expressed as,
\[
\hat{U}=\frac{1}{n^2}\sum_{i, j}\hat{K}_{i,j}\hat{S}_{i,j}.
\]
We include the diagonal terms in the covariate-adjusted similarities because they also
contain the similarity information after the adjustment. In fact, the covariate-adjusted
GSU is a weighted V statistic, and its asymptotic distribution can be attained similarly as weighted U statistic. We use matrix eigen-decomposition to approximate
the eigen-values in function decomposition. Let $\{\hat{\lambda}_{s}\}$
and $\{\hat{\eta}_{t}\}$ be the eigen-values for matrices
$\hat{K}$ and $\hat{S}$ respectively,
the limiting distribution of $U$ is given by (Supplementary Appendix S7),
\[
n\hat{U}\sim\frac{1}{n(n-P-1)}\sum_{t=1}^{n} \hat{\eta}_{t} \sum_{s=1}^{n} \hat{\lambda}_{s}\chi_{st}^{2},
\]
where $\{\chi_{st}^{2}\}$ are independent chi-square random variables
with 1 degree of freedom. The p-value can be calculated by using the
Davies' method \citep{Davies1980}, the Liu's method\citep{Liu2009a}
or the Kuonen's method\citep{Kuonen1999}. To facilitate the high dimensional data analysis, we developed a C++ package based on GSU (https://github.com/changshuaiwei/gsu).

\section{Simulation study}

\subsection{Simulation method}

To mimic real genetic structure, we used genetic data from the 1000
Genome Project\citep{Abecasis2010}. Based on the genetic data, we
then simulated phenotype values. In particular, we used a 1Mb region
of the genome (Chromosome 17: 7344328-8344327)
from the 1000 Genome Project. For each simulation replicate, we randomly
chose a 30kb segment from the 1Mb region and formed a SNV-set for
the analysis, in which only rare variants (i.e., $MAF<0.05$) are used except otherwise specified. From the SNV-set, we
set a proportion of the SNVs as causal. A number of individuals were
randomly chosen from the total 1092 individuals as the simulation sample to study the performance
of the methods. We set sample size $n=50$ by default. 

To investigate the robustness
against different phenotype distributions, we simulated four types
of phenotypes:
\begin{enumerate}
\item A binary-distributed phenotype (denoted as B), by $logit(P(Y_{i}=1))=\mu_i+G_{i}^{T}\beta$,
\item A Poisson-distributed phenotype (denoted as P), by $Y_i \sim Pois(a_i), log(a_i)=\mu_i + G_{i}^{T}\beta$,
\item A Gaussian-distributed phenotype (denoted as G), by $Y_{i}=\mu_i+G_{i}^{T}\beta+\varepsilon_{i},\:\varepsilon_{i}\sim N(0,\sigma^{2})$,
\item And a Cauchy-distributed phenotype (denoted as C), by $Y_{i}\sim cauchy(a_{i},b),\: a_{i}=\mu_i+G_{i}^{T}\beta$,
\end{enumerate}
Here, $Y_{i}$ and $G_{i}$ were the phenotype value and the genotype
vector (coded as 0, 1, and 2) for the $i$-th individual, respectively. We set $\mu_i=0$ except otherwise specified.
$\beta$ were the effects of the SNVs, which were sampled from a uniform
distribution with a mean of $\mu_{\beta}$ and a variance of $\sigma_{\beta}^{2}$.

Three sets of simulations were performed. In simulation I, we considered
a single phenotype; in simulation II, we considered multivariate phenotype; in simulation III, we considered multivariate phenotype under the influence of confounding effects. Details of simulation settings are in Supplementary Appendix S8.

We evaluated the performance of GSU by comparing it with variance component score (VCscore) test under univariate or multivariate linear mixed model \citep{Wu2011, Maity2012}. For
each simulation, we created 1000 simulation replicates to evaluate
type I error and power. Type I error rates and powers are calculated using percentage of p-values smaller than a given threshold (e.g., 0.05) under null models and alternative models respectively.

\subsection{\textit{Result for Simulation I}}

The type I error rates and powers are summarized in Table \ref{Tab:1}.
GSU had a well-controlled type I error (around 0.05) for all 4 phenotypes, while VCscore had an inflated
type I error rates (0.113) for Cauchy-distributed
phenotype and over-conservative type I error rates (0.005) for Binary-distributed phenotype. 

For the disease model where half of the causal SNVs were deleterious
(Table \ref{Tab:1}), GSU had slightly lower power than VCscore for Gaussian-distributed (0.258 v.s. 0.345) and Poisson-distributed phenotype (0.506 v.s. 0.651), but had significantly higher power than VCscore for Cauchy-distributed (0.503 v.s. 0.21) and Binary-distributed phenotype(0.402 v.s. 0.083). The same comparison was observed for the second disease model in which a majority of the SNVs were deleterious. 

We performed additional simulations by including both common and rare variants (Supplementary Table S4). Under this setting, the power of VCscore increased significantly for Binary phenotype (0.764), though still lower than that of GSU (0.807). GSU attained higher power than VCscore for Poisson (0.813 v.s. 0.795) and Cauchy (0.885 v.s. 0.573) phenotype. Nevertheless, GSU was still less powerful than VCscore for Gaussian phenotype (0.853 v.s. 0.878).

\begin{table}[htbp]
\centering 
\caption{Type I errors and Powers for the univariate analysis}
\label{Tab:1}
\begin{minipage}{85mm}
\begin{tabular}{cccccc}
\hline
Model\footnote{Alt$_1$ represents settings of $\mu_{\beta}=0$
and $\sigma_{\beta}^{2}>0$; Alt$_2$ represents settings of $\mu_{\beta}>0$
and $\sigma_{\beta}^{2}>0$.}  & Method  & \multicolumn{4}{c}{Distribution\footnote{B, C, G, P represent Binary-distributed, Cauchy-distributed, Gaussian-distributed,
and Poisson-distributed phenotypes, respectively.} }\tabularnewline
\hline
 &  & B & C & G & P\tabularnewline
\hline
Null & Vcscore & 0.005 & 0.113 & 0.019 & 0.047\tabularnewline
 & GSU & 0.044 & 0.051 & 0.047 & 0.058\tabularnewline
\hline
Alt$_1$ & Vcscore & 0.083 & 0.21 & 0.345 & 0.651\tabularnewline
 & GSU & 0.402 & 0.503 & 0.258 & 0.506\tabularnewline
\hline
Alt$_2$ & Vcscore & 0.023 & 0.434 & 0.747 & 0.942\tabularnewline
 & GSU & 0.458 & 0.753 & 0.628 & 0.864\tabularnewline
\hline
\end{tabular}
\end{minipage}
%\vspace*{6pt}
\end{table}

\begin{table}[htbp]
\centering 
\caption{Type I errors and Powers for the multivariate analysis}
\label{Tab:2}
\begin{minipage}{85mm}
\begin{tabular}{cccccc}
\hline
Model  & Method  & \multicolumn{4}{c}{Distribution\footnote{B, C, G, P represent Binary-distributed, Cauchy-distributed, Gaussian-distributed,
and Poisson-distributed phenotypes, respectively.} }\tabularnewline
\hline
 &  & BPP & CGG & BBG & BCG\tabularnewline
\hline
Null & Vcscore & 0.054 & 0.194 & 0.049 & 0.179\tabularnewline
 & GSU & 0.051 & 0.043 & 0.049 & 0.055\tabularnewline
\hline
Alt & Vcscore & 0.939 & 0.273 & 0.478 & 0.309\tabularnewline
 & GSU & 0.84 & 0.664 & 0.716 & 0.684\tabularnewline
\hline
\end{tabular}
\end{minipage}
\end{table}

\begin{table}[htbp]
\centering 
\caption{Type I errors at different significance levels}

\label{Tab:3}
\begin{minipage}{85mm}
\begin{tabular}{cccccc}
\hline
Level & Method  & \multicolumn{4}{c}{Distribution\footnote{B, C, G, P represent Binary-distributed, Cauchy-distributed, Gaussian-distributed,
and Poisson-distributed phenotypes, respectively.} }\tabularnewline
\hline
& & BPP & CGG & BBG & BCG\tabularnewline
\hline
$1 \times 10^{-2}$ &Vcscore & 0.017 & 0.133 & 0.013 & 0.138\tabularnewline
&GSU & 0.013 & 0.015 & 0.011 & 0.015\tabularnewline
\hline
$5 \times 10^{-3}$ &Vcscore & 0.011 & 0.120 & 0.0078 & 0.124\tabularnewline
&GSU & 0.0074 & 0.0091 & 0.0059 & 0.0093\tabularnewline
\hline
\end{tabular}
\end{minipage}
\end{table}

\subsection{\textit{Result for Simulation II}}
The type I error rates and powers for the multivariate analysis are summarized
in Table \ref{Tab:2}. Similar to the results of the univariate analysis,
GSU can correctly control type I error at the level of 0.05 (Table
\ref{Tab:2}), while VCscore had inflated type I error
when the phenotype contained variables with heavy tailed distribution (e.g., CGG and BCG). GSU attained higher power than VCscore for BBG, CGG and BCG phenotypes, and similar power as VCscore for BPP phenotype.

We examined the type I error rates at more stringent significance levels (Table \ref{Tab:3}) by simulating 1 million replicates. In general, GSU can control the type I error better than VCscore. For example, at $5 \times 10^{-3}$ and for BPP phenotype, GSU had type I error near $5 \times 10^{-3}$ (i.e., $7.4 \times 10^{-3}$), while, VCscore had type I error much higher than $5 \times 10^{-3}$ (i.e., $1.1 \times 10^{-2}$). While simulation demonstrated robustness of GSU over VCscore on controlling type I error, we observe GSU has slightly inflated type I errors at $5 \times 10^{-3}$ level. We suspect this is because of the small sample size.  We therefore conducted another set of simulation with sample size of 200, and the results showed type I errors of GSU are better controlled for stringent significant levels under larger sample size (Supplementary Table S5). 

To separate influences of different distributions, we also compared GSU and VCscore when phenotype have the same distributions (i.e., BBB, CCC, GGG, PPP). The results (Supplementary Table S6) are similar to those for univariate phenotype. In general, GSU can control type I errors better than VCscore. GSU had slightly lower power than VCscore for GGG phenotype (0.882 v.s. 0.958) and PPP phenotype (0.862 v.s. 0.966), but attained significantly higher power for BBB phenotype (0.862 v.s. 0.26) and CCC phenotype (0.724 v.s. 0.284). We further increased the dimension of phenotype to 10 for each type, and the comparisons showed that GSU have better control of type I error and attain higher power for most cases (Supplementary Figure S1).

\begin{table}[htbp]
\centering 
\caption{Type I errors for multivariate analysis with moderate confounding effects}

\label{Tab:4}
\begin{minipage}{85mm}
\begin{tabular}{cccccc}
\hline
Adj\footnote{Adj represents whether covariate adjustments are performed.}  & Method  & \multicolumn{4}{c}{Distribution\footnote{B, C, G, P represent Binary-distributed, Cauchy-distributed, Gaussian-distributed,
and Poisson-distributed phenotypes, respectively.} }\tabularnewline
\hline
 &  & BPP & CGG & BBG & BCG\tabularnewline
\hline
Yes & Vcscore & 0.322 & 0.174 & 0.054 & 0.184\tabularnewline
 & GSU & 0.061 & 0.056 & 0.056 & 0.057\tabularnewline
\hline
No & Vcscore & 0.408 & 0.171 & 0.217 & 0.183\tabularnewline
 & GSU & 0.135 & 0.113 & 0.147 & 0.115\tabularnewline
\hline
\end{tabular}
\end{minipage}
\end{table}

\subsection{\textit{Result for Simulation III}}

We summarized the type I errors in Table \ref{Tab:4}. Without covariates adjustment, both methods had inflated type I errors. With covariates adjustment, GSU showed robustness against confounding effects for all 4 multivariate phenotypes, with type I errors ranging from 0.056 to 0.061. VCscore can control type I error for BBG phenotype (0.054), but had inflated type I errors, ranging from 0.174 to 0.322, for the other 3 multivariate phenotypes.

\begin{figure*}[htbp]

\centerline{\includegraphics[scale=0.6]{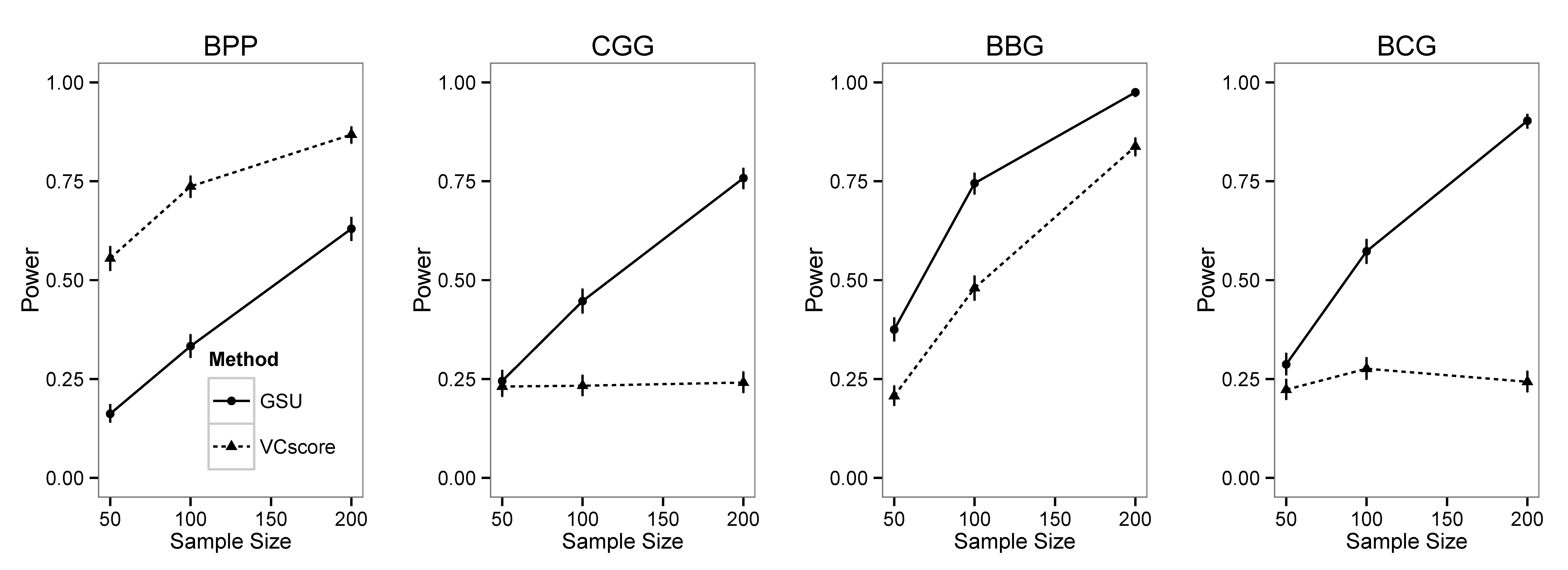}}
\caption{Power comparison for the multivariate analysis after adjusting for confounding effects}

\label{Fig:1} 
\end{figure*}

In Figure \ref{Fig:1}, we generated the power curves by plotting the powers of the two methods against different sample sizes (50 to 200). GSU has higher power than VCscore for different sample sizes and multivariate phenotypes, except for BPP phenotype. The ``higher power" of VCscore for BPP phenotype is due to the fact that VCscore has inflated type I error (i.e., 0.322, as shown in Table \ref{Tab:4}).

\section{Real Data Application}

We analyzed the whole genome sequencing data (WGS) from Alzherimer's Disease Neuroimaging Initiative (ADNI) using the GSU C++ package. ADNI is a large scale longitudinal study that collects and utilize various predictors of Alzherimer's Disease, including 3D brain imaging, cognitive measurements and genetic data. The sample with WGS data contains 808 individuals, with 280 Normal Controls (NC), 234 Early Mild Cognitive Impaired patients (EMCI), 246 Late Cognitive Impaired patients (LMCI), and 48 Alzheimer's Disease patients (AD) at study baseline. 

Whole genome sequencing was performed on autosomal chromosomes for each subject. To form SNV-set, we group the genetic variants based on the gene range list from GRch37 assembly, where we only used the non-overlapping genes. For genetic variants outside of gene ranges, we group them by evenly spacing the remaining genome with windows of 50kb. After completing quality control(e.g., delete variants with high missing rate) and grouping process, about 21 millions genetic variants remained for analysis, forming 61683 SNV-sets. 

We were interested in testing the association
of the SNV-sets with brain imaging summary matrices considered important to cognitive impairment. In particular, we used 6 variables: 18F-fluoro-2-deoxyglucose (FDG), Hippocampus,	Entorhinal, 8F-florbetapir (AV45),	Fusiform, and Ventricles measurements at base-line, as multivariate phenotype. The phenotype similarity is calculated using weighted Laplacian kernel, $S_{i,j}^{LK}=exp(-\sum_{l=1}^{L}\omega_{l}|y_{i,l}-y_{j,l}|)$. We ``fished" the weight $\omega_{l}$ from the case control status. In particular, we regressed the case control status on the scaled multivariate phenotype and obtained regression coefficient $\beta_l$ for $l$-th variable, where we assigned $\omega_l=|\beta_l|$ (Table S7). 

In order to adjust the potential confounding effects, we included age, gender, race and top 20 genome principle components as covariates in the analysis. Two sets of whole genome association analysis were performed. For the first scan, we include both common and rare variants, while for the second scan we only include rare variants. The QQ plots (Figure S2 and Figure S3) showed no systematical bias after adjusting covariates. We listed the top 5 SNV-sets for each scan in Table \ref{Tab:5}. When both common and rare variants were considered, 4 SNV-sets (i.e., \textit{APOE}, Ch19-45389309-45439308, \textit{APOC1}, \textit{TOMM40}) pasted the Bonferroni threshold, among which the genes \textit{APOE} and \textit{TOMM40} has been reported in previous studies. As a comparison, we also performed the analysis using VCscore (Supplementary Table S10). VCscore attained similar results for the top association findings, though with less significant p-values (e.g., p-value =$1.98\times10^{-26}$ for \textit{APOE}). 

When only rare variants are considered, no SNV-set past the Bonferroni threshold. Interestingly, the gene \textit{APOC1} was listed as one of the top 5 associated genes from both analyses. Further investigation will be needed to study its role in AD. More detailed results are in Table S8 and S9. We further calculated the p-value of the top SNV-sets using AD case-control status instead of multivariate phenotype with 6 intermediate measurements. The univariate analysis attained less significant result (Table S11). For example, the p-value of \textit{APOE} is $3.44\times 10^{-8}$ from analysis using AD case control status, less significant than $2.77\times 10^{-48}$ from analysis using brain imaging matrices.

\begin{table}[htbp]
\centering 

\caption{Top SNV-sets in the multivariate analysis of whole genome sequencing data in ADNI study}

\begin{minipage}{85mm}

\label{Tab:5}

\begin{tabular}{cccc}
\Hline
{SNV-set\footnote{SNV-set is named either using gene name, or with the format of ``chromosome - starting position - ending position", where the position is referred to GRch37 assembly.}} & Chr & {Size\footnote{Number of SNV in the SNV-set}} & p-value\tabularnewline
\hline
\multicolumn{4}{c} {Common and Rare Variants} \tabularnewline
\hline
\textit{APOE} & 19 & 17 & $2.77\times 10^{-48}$\tabularnewline
Ch19-45389309-45439308 & 19 & 162 & $1.64\times 10^{-37}$\tabularnewline
\textit{APOC1} & 19 & 37 & $3.38\times 10^{-31}$\tabularnewline
\textit{TOMM40} & 19 & 126 & $9.28\times 10^{-19}$\tabularnewline
\textit{RHPN2} & 19 & 758 & $2.54\times 10^{-06}$\tabularnewline
\hline
\multicolumn{4}{c} {Rare Variants} \tabularnewline
\hline
Ch1-107013494-107063493 & 1 & 240 & $1.96\times 10^{-06}$\tabularnewline
\textit{APOC1} & 19 & 26 & $6.28\times 10^{-06}$\tabularnewline
Ch17-40300052-40350051 & 17 & 64 & $2.51\times 10^{-05}$\tabularnewline
Ch4-189560456-189610455 & 4 & 314 & $3.14\times 10^{-05}$\tabularnewline
\textit{LOC101927616} & 12 & 107 & $3.16\times 10^{-05}$\tabularnewline
\hline
\end{tabular} 
\end{minipage}
\end{table}

\section{Discussion} 

Many genetic studies collect multiple secondary phenotypes,
or use intermediate biomarkers, to study complex diseases. By considering
multiple phenotypes that measure the different aspects of underlying
diseases, the power of the association analysis can potentially be improved
\citep{Zhang2010,Maity2012}. Several methods
were recently developed to detect the joint effect of genetic variants on multivariate phenotype\citep{Tao2015,wang2015}. Most were built on parametric framework that poses certain assumptions on phenotype distribution. In this paper, we proposed a non-parametric test, GSU, based on similarity measurement. Simulation study showed that our methods can can control type I error for multiple different phenotypes and moderate level of confounding effects. In most cases, GSU also attained higher power than the parametric method. Although the simulation results depend on
the simulation settings, and should always be interpreted in the context
of the simulation setting, we believe the results reflect the advantage
of GSU in a broader sense, because 1) the genetic data used in the
simulation comes from the 1000 Genome Project, which reflects the
LD pattern and the allele frequency distribution in the general
population; and 2) we simulated a wide range of disease models, including univariate phenotype and multivariate phenotype with different distributions,  to
mimic real disease scenarios. 

The test statistics in VCscore is a quadratic form, $T=\hat{Y}^TK\hat{Y}$, where $\hat{Y}$ is the standardized residual under null, and $K$ is the genetic similarity matrix. If we rewrite $T$ as $T=\sum_{i,j} K_{i,j} (\hat{Y}_i\hat{Y_j})$, VCscore is actually a weighted V statistic with cross product kernel $\hat{S}_{i,j}=\hat{Y}_i\hat{Y_j}$. In this respect, VCscore can be considered as a special case of GSU. Nonetheless, there are several key differences: 1) GSU allows general forms of similarity and thus can be used for association analysis of elements in general metric space; 2) For multivariate association analysis, GSU with LK based similarity has the ability to detect any types of association (strongly positive definite similarity) and its asymptotic test is robust against distribution assumptions (bounded similarity); 3) For covariates adjustment, GSU used a centralized similarity $\tilde{S}_0$ and then perform two sided projection, i.e., $\hat{S}=(I-X(X^TX)^{-1}X^T)\tilde{S}_0(I-X(X^TX)^{-1}X^T)$, while, VCscore performed two sided projection on original similarity ($S=YY^T$), i.e., $\hat{S}=(I-X(X^TX)^{-1}X^T)S(I-X(X^TX)^{-1}X^T)$; 4) Asymptotic distribution of GSU is in the form of $\sum_{t=1}^{\infty}\eta_{t}\sum_{s=1}^{\infty}\lambda_{s}\chi_{st}^{2}$, where distribution of VCscore is in the form of $\sum_{s}\lambda_{s}\chi_{s}^{2}$; 5) For multivariate phenotype with $L$ variables, the dimension for similarity matrix is $n\times n$ in GSU and $nL\times nL$ in VCscore. 

In simulation studies, we observed higher power of GSU over VCscore. This is mainly due to the fact that GSU is equipped with strongly positive definite kernel which can detect any type of association while the cross product kernel in VCscore does not have this property. We performed another set of simulations by generating dependence structure via rotation operator (Supplementary Appendix S9). In particular, we first generate two i.i.d. multimodal continuously distributed variables and then rotate the vector with angle $\theta \in (0,\pi/4)$ (Figure S4). The data generated thus does not have first order dependence structure (correlation) nor second order dependence structure. The result (Figure S5) showed that GSU ( with LK-based similarity) had power of 1 for large enough sample size, while VCscore (with cross product kernel) can not detect any association regardless of different sample sizes. Though the``toy" simulation may not represent common scenarios in genetic association studies, it empirically explains the reason why GSU attained higher power than VCscore. To further investigate the influence of different kernels, we performed simulations using 5 different kernels for GSU, including 3 strongly positive definite kernels. The result shows that GSU with strongly positive definite kernels have higher powers for the most of the time, among which GSU with LK kernel have highest power (Supplementary Figure S6). In general, we recommend to use LK kernel for GSU. Nevertheless, its performance may not guaranteed to be optimal. In this case, we can perform kernel selection, for example, by using the procedure proposed by \cite{Wu2013}. Besides the choice of kernel, different choices of weights can also influence the power of GSU for multivariate phenotype. In principle, we should use weights that represent their relative importance with respect to  the underlying "true phenotype". For example, in real data analysis, we obtained the weights based on their contributions to the AD disease status. Here in this paper, we only considered the joint effect of SNV-sets. If gene environment interaction effects are to be considered, we can calculate a composite similarity using both the genetic information and environmental information \citep{Wei2016, tong2016genome}, and then construct GSU with the composite similarity and the phenotype similarity.

The asymptotic test for GSU (with LK-based similarity) is shown to be robust to distribution assumption. This is because the Laplacian kernel is bounded between 0 and 1, and the resulting similarities $h(\cdot,\cdot)$ and $f(\cdot,\cdot)$ thus satisfy the regularity condition of asymptotic test, i.e., $E(h(Y,Y))<\infty$ and $E(f(G,G))<\infty$. However, cross-product kernel does not have this property. As a result, we observed that in simulation studies GSU had more robust type I errors than VCscore. Nevertheless, we still observed slightly inflated type I error with stringent significant level (e.g., $5\times 10^{-3}$) when $n=50$. This is because the asymptotic null distribution can not approximate the actually null distribution well when sample size is small compared with when sample size is large (Supplementary Figure S7). One way to improve the robustness for small sample size is to take an rank transformation for each variable (i.e., $r_{i,j}=( rank(y_{i,j}) -0.5) /n$) before calculating the similarity. We performed additional simulation for GSU with rank transformation for $n=50$ using same setting as simulation II. The results showed that GSU with rank transformation (GSU-rk) can control type I error well even with more stringent significant level for small sample size (Supplementary Table S12). Nevertheless, rank transformation can cause loss of information, which might lead to lower power.

In simulations, we observe that VCscore, although designed for Gaussian distributed phenotype, appears to be able to control type I error appropriately and attain slightly higher power for Poisson phenotype. This may be due to that Poisson distribution can be reasonably approximated by Gaussian distribution when its mean is moderate to large. We performed additional simulation using heavily right skewed Poisson distribution, and the results showed VCscore had lower power for one simulation and inflated type I errors for another simulation (Supplementary Table S13). We can use rank transformation to improve the robustness of VCscore \citep{Wei2016}. We performed additional simulation to compare GSU-rk to VCscore test with rank transformation (VCscore-rk). The result (Supplementary Table S14) showed that VCscore-rk can control Type I errors under various setting. However, VCscore-rk still had lower powers than GSU-rk for most cases. 

For the analysis of multivariate phenotype, the difference on the dimension of similarity matrix for GSU and VCscore influenced the computation efficiency especially when the number of variables in multivariate phenotype increases. The key reason is the cost of the eigen decomposition. For analysis of $L$-variable multivariate phenotype in a sample of size $n$, GSU  needs to decompose a $n\times n$ matrix, while VCscore needs to decompose a $Ln \times Ln$ matrix. The time used for matrix decomposition are $O(n^3)$ for GSU and $O(L^3n^3)$ for VCscore. For example, in real data application when $L=6$, the average time to analyze one SNV set is 36.75 seconds for VCscore and 1.3 seconds for GSU. For high-dimensional setting (e.g., $L>>n$), VCscore is computationally infeasible. An additional simulation shows that GSU is well behaved when the dimension of phenotype increase to 100 (Supplementary Figure S8). Nevertheless, noises in high dimensional phenotype or genotype may reduce the power of GSU. In this case, dimension reduction techniques, such as variable selection and principle component analysis, can be used to increase power.

The covariate adjustment proposed in the paper is a heuristic approach for adjusting confounding effect. Accurate adjustment of confounding effects requires additional assumptions on the distributions and the functional forms between responses and covariates. In the paper, we showed GSU works well when the confounding effects are moderate. Nonetheless, the heuristic covariate adjustment in GSU should always be used with caution. If there is a strong confounding effect, the heuristic approach might not control type I error very well. For this paper, covariate adjustment is not the primary focus, and the issue will be investigated in future studies. 

Besides confounding effects, the correlation among variables in multivariate phenotype may also influence the performance of association testing. This is particularly important for regression based methods, since it handles multivariate phenotype by stretching the phenotype matrix to a long phenotype vector. Without considering correlations among variables in phenotype, the test will lead to inflated type I error. Nevertheless, GSU don't have this issue, since its similarity matrix is calculated on subject level and its inference only assume independence between subjects. We performed additional simulations by introducing additional correlation in the multivariate phenotype (Supplementary Table S15). The results showed that, in general, GSU can control type I error and attain higher power than VCscore (Supplementary Figure S9).  

In recent years, U-statistic-based methods became popular
in genetic data analysis, and have shown their robustness and flexibility
for analyzing genetic data\citep{Schaid2005,li2011,Wei2015,Wei2016}.
GSU is a general framework of association analysis and is based on similarity measurements and U statistics. 
In this paper, we have focused on the association analysis between multivariate phenotype and categorical sequencing
data (i.e, SNV data). GSU can easily be applied to analyze other types of genetic data,
such as count data (CNV data) and continuous data (expression data) with unified LK-based similarity (Section 3.2). With appropriate similarity measurement (Section 2.1), GSU can also be used for association testing of modern data types, such as imaging, curves and trees.

\backmatter

%%%%%% include this section if you wish to acknowledge people,
%%%%%% grant support, etc.

\section*{Acknowledgements}
Data used in preparation of this article were obtained from the Alzheimer Disease Neuroimaging Initiative (ADNI) database
(adni.loni.usc.edu). As such, the investigators within the ADNI contributed to the design and implementation of ADNI and/or
provided data but did not participate in analysis or writing of this report. A complete listing of ADNI investigators can be
found at: http://adni.loni.usc.edu/wp-content/uploads/how{\_}to{\_}apply/ADNI{\_}Acknowledgment{\_}List.pdf

%The authors thank XXX .\vspace*{-8pt}

%%%%%% include this section only if your manuscript refers to supplementary
%%%%%% materials -- see Instructions for Authors at 
%%%%%% http://www.tibs.org/biometrics

\section*{Supplementary Materials}

Supplementary Materials are available on line.

\bibliographystyle{biom} \bibliography{multi-trait}

\appendix
\label{appendix}

%\section{}
%\section*{Appendices: Proof and Computation Details}

\section*{Appendix 1}
\label{appendix}
Due to space limit, we here only sketch the proofs. Detailed proofs can be found at Supplementary Appendix S2 and S3.
\subsection*{Appendix A: embedding into Hilbert Space}
\label{sub:Proof-of-independence}

For each positive definite kernel $h$, we can construct a unique reproducing kernel Hilbert space (RKHS) $\mathcal{H}$ with reproducing kernel $h$ \citep{Berlinet2011}, such that, 1) $\forall y \in \Psi_Y$, $h(\cdot, y) \in \mathcal{H}$, 2) $\forall y \in \Psi_Y$, $\forall \varphi \in \mathcal{H}$, $<\varphi, h(\cdot, y)>_{\mathcal{H}}=\varphi(y)$. We can write $h(\cdot,y)=\tau_h(y)$, and then represent a measure $\vartheta \in \mathcal{M}$ as an element in RKHS \citep{Lyons2013} using an embedding map $\pi$: $\mathcal{M} \to \mathcal{H}$, s.t.,$ \pi_h(\vartheta)=\int \tau_h(y) d\vartheta(y)=\int h(.,y) d\vartheta(y) 
$. Further, if $h$ is strongly positive definite, we can show the mapping $\pi_h$ is one-to-one, i.e., $\vartheta_1=\vartheta_2 \Leftrightarrow \pi(\vartheta_1)= \pi(\vartheta_2)$.

Let $\tau_h(y)=h(\cdot,y)$ and $\tau_f(g)=f(\cdot, g)$. We can then write $\mu_U$ as, $\mu_U=|| \pi_{f \otimes h} (\vartheta)||_{\mathcal{H}}^2$, where, $ \pi_{f \otimes h} (\vartheta) = \int\int \tau_f(g)\tau_h(y) d \vartheta(g,y)$. 
If $\mu_U=0$, then we know $\pi_{f \otimes h} (\vartheta)=0$, i.e.,
\[
\int f(g_1,g)h(y_1,y)d \vartheta(g_1,y_1)=0, \text{ } \forall (g,y) \in \Psi_G \times \Psi_Y.
\]
We then can show, by repeatedly using measure embedding, that $\forall A \subset \Psi_G, \forall B \subset \Psi_Y, \int 1_A(g_1) 1_B(y_1) d \vartheta(g_1,y_1) = 0$, i.e., $G\Perp Y$.

\subsection*{Appendix B: Proof of Theorem 3}
\label{sub:Proof-of-Thorem-1}

Because of the orthogonality of $\{\phi_{s}(\cdotp)\}$ and the fact that $E(\tilde{h}(Y_{1},Y_{2})|Y_{1})=0$, we can show
$
E\phi_{s}(Y)=0,  \forall\, s>1
$. Similarly, $E\varphi_{t}(G)=0,  \forall\, t>1$. Under the null hypothesis, predictor element ($G_{i}$) is independent of response element
($Y_{i}$). Therefore, for $s>1$ and $t>1$,
\[
E(\eta_{t}^{\star}(G_{1})\phi_{s}^{\star}(Y_{1}))
=\eta_{t}^{0.5}E\varphi_{t}(G_{1})\lambda_{s}^{0.5}E\phi_{s}(Y_{1})
=0,
\]
and
\[
E(\eta_{t}^{\star}(G_{1})\phi_{s}^{\star}(Y_{1})\eta_{t'}^{\star}(G_{1})\phi_{s'}^{\star}(Y_{1}))
= \begin{cases}
\eta_{t}\lambda_{s}, & \text{if \ensuremath{s=s'}and \ensuremath{t=t'}}\\
0, & \text{otherwise.}
\end{cases}
\]
Therefore, for any finite subset $\Delta$ of $\{(s,t)\}_{s>1,t>1}$, the multivariate random variable
$\left\{ \frac{1}{\sqrt{n}}\sum_{i=1}^{n}\eta_{t}^{\star}(G_{i})\phi_{s}^{\star}(Y_{i})\right\} _{(s,t)\in\Delta}$
converges to a multivariate normal distribution.

Then, we need to show the convergence is uniform. Notice that,
$
\sum_{s>1,t>1}E(\eta_{t}^{\star}(G_{1})\phi_{s}^{\star}(Y_{1}))^{2}
= E(h(Y,Y))E(f(G,G))< \infty
$. Under the condition $\sum_{s>1,t>1}E(\eta_{t}^{\star}(G_{1})\phi_{s}^{\star}(Y_{1}))^{2}<\infty$,
the infinite countable sequence of function $\{\eta_{t}^{\star}(\cdotp)\phi_{s}^{\star}(\cdotp)\}$
is a Donsker class (Theorem 2.13.1 in \cite{Vaart2000}). Therefore, the empirical process,
$\frac{1}{\sqrt{n}}\sum_{i=1}^{n}\eta_{t}^{\star}(G_{i})\phi_{s}^{\star}(Y_{i})$,
converges weakly to the Gaussian process $Z_{s,t}$ with mean zero
and covariance function, $
cov(Z_{s,t},Z_{s',t'})=E(\eta_{t}^{\star}(G_{1})\phi_{s}^{\star}(Y_{1})\eta_{t'}^{\star}(G_{1})\phi_{s'}^{\star}(Y_{1}))$.
With this uniform convergence (for all $s>1$ and $t>1$), we can
show that,
\[
nU \xrightarrow{D} \sum_{t=2}^{\infty}\sum_{s=2}^{\infty}(Z_{s,t})^{2} -\sum_{t=2}^{\infty}\sum_{s=2}^{\infty}\eta_{t}\lambda_{s}=\sum_{t=1}^{\infty}\eta_{t}\sum_{s=1}^{\infty}\lambda_{s}(\chi_{st}^{2}-1),
\]
where $\chi_{st}^{2}$ are i.i.d chi-squared random variables with
a d.f. of 1.

\label{lastpage}

\end{document}